# Dynamic Magnetoelectric Effects in Bulk and Layered composites of Cobalt Zinc Ferrite and Lead Zirconate Titanate

G. Srinivasan, R. Hayes, and C. P. DeVreugd
Physics Department, Oakland University, Rochester, MI 48309, USA

V. M. Laletsin and N. Padubnaya
Institute of technical acoustics, National Academy of Sciences, 210717 Vitebsk, Belarus

ABSTRACT

Low frequency magnetoelectric (ME) coupling has been investigated in bulk and multilayers of cobalt zinc ferrite, $Co_{1-x}Zn_xFe_2O_4$ (x=0-0.6), and lead zirconate titanate. In bulk samples, the transverse and longitudinal couplings are weak and are of equal magnitude. A substantial strengthening of ME interactions is evident in layered structures, with the ME voltage coefficient a factor of 10-30 higher than in bulk samples. Important findings of the studies in layered composites are as follows. (i) The transverse coupling is stronger than the longitudinal coupling. (ii) The strength of ME interactions is dependent on Zn substitution with a maximum for x=0.4. (iii) Analysis of volume and static magnetic field dependence of ME voltage coefficients reveal a weak coupling at the ferromagnetic-piezoelectric interface. (iv) The interface coupling k increases with Zn substitution and k versus x profile shows a maximum centered at x=0.4. (iv) The Zn-assisted enhancement is attributed to efficient magneto-mechanical coupling in the ferrite.

## 1. Introduction

In a magnetoelectric (ME) material, an applied magnetic field produces a dielectric polarization or an external electric field results in an induced magnetization [1]. The induced polarization **P** is related to the magnetic field **H** by $\mathbf{P} = \alpha \mathbf{H}$, where $\alpha$ is the ME-susceptibility. For a single-phase material to be magnetoelectric, one requires a long-range magnetic order and the presence of permanent electric dipoles. The (static) effect, first observed in antiferromagnetic $Cr_2O_3$, is weak in single-phase compounds [2].

But a composite of piezomagnetic-piezoelectric phases is also expected to be magnetoelectric since $\alpha = \delta P/\delta H$ is a "product-property" that results from magnetostriction induced deformation and piezoelectric charge generation [3-5]. We are interested in the dynamic ME effect; for an ac magnetic field $\delta H$ applied to a biased composite, one measures the induced voltage $\delta V$. The ME voltage coefficient $\alpha'_E = \delta V/t'\delta H$ and is related to the ME susceptibility through $\alpha = \varepsilon_o \varepsilon_r \alpha'_E$ where $t'$ is the composite thickness and $\varepsilon_r$ is the relative permittivity.

This report is concerned with the synthesis of bulk and layered composites of ferromagnetic-ferroelectric oxides and studies on the nature of magnetoelectric interactions. Composites of interest in the past were bulk samples of nickel ferrite or cobalt ferrite with $BaTiO_3$ or lead zirconate titanate (PZT) [3,4]. Powders of the two oxides are mixed and sintered. Such bulk samples in general show weak ME effects. The main cause is low resistivity for ferrites that gives rise to a leakage current and limits the electric field for orienting the dipoles, leading to poor piezoelectric coupling and loss of induced charges. These problems could be eliminated in a layered structure [4]. Series electrical connectivity gives rise to a high resistivity, negligible leakage current and strong piezoelectric and ME coupling. Significant developments in this regard are the recent observation of giant ME effects in bilayers and multilayers of lanthanum manganite-PZT and nickel ferrite-PZT [5-8].

This work is directed toward an understanding of the effects of magnetic parameters of ferrites on ME interactions in bulk and multilayers of $Co_{1-x}Zn_xFe_2O_4$ (CZFO) (x=0-0.6) and PZT. It is possible to accomplish controlled variations in magnetic parameter with Zn substitution in ferrites [7]. We prepared bulk samples by mixing and sintering powders for the two phases. Layered samples were prepared using thick films obtained by tape casting. X-ray diffraction studies indicate samples free of impurities, but a strained PZT lattice in layered samples. Magnetic and electrical parameters for the composites were in general agreement with bulk values.

Bulk samples show weak ME interactions. We find a substantial enhancement in the strength of ME interactions in layered samples. Further improvement in ME coupling in layered samples is accomplished with the introduction of Zn in the ferrite. The ME voltage coefficient increases as the Zn concentration is increased and shows a maximum for x = 0.4. The coupling weakens for higher Zn substitution. These data have been analyzed using our recent model for a bilayer of ferrite-PZT [9,10]. A unique feature in the model is the introduction of an interface coupling parameter k, with k=1 for an ideal interface and k=0 for the frictionless case. For CZFO-PZT, k values range from 0.1 to 0.6 with a maximum for x=0.4. The results are discussed in terms of k dependence on structural, mechanical, chemical, and electromagnetic parameters for the two phases. We infer from our analysis that enhancement in ME coupling with Zn substitution is related to a reduction in magnetic anisotropy, leading to high permeability and, therefore, a strong magnetomechanical effect in the ferrite. The multilayer ferrite-PZT structures studied here show a strong ME coupling and are of interest for use as sensors and actuators.

## 2. Experiment

Bulk composites of cobalt zinc ferrite-PZT were prepared with submicron powders of CZFO and commercial PZT (American Piezoceramics- APC 850). Samples with PZT amounts varying for 0 to 100 vol.% were made. Ferrite and PZT were mixed in a ballmill, pressed into pellets and sintered at 1300-1450 K for 1 hr in a traditional furnace. The two phases cosinter well and do not form impurity phases for sintering temperatures below 1500 K [7]. For microwave sintering, samples were placed in a susceptor and heated at 1200 K for 30 min. in a 1800-W furnace operating at 2.48 GHz.

Layered samples were made using thick films of ferrite and PZT obtained by tape casting. For tape casting, powders of ferrite or PZT were mixed with a solvent (ethly alcohol) and a dispersant and ball milled for 24 hrs, followed by a second ball milling with a plasticizer (butyl benzyl phthalate) and a binder (polyvinyl butyral) for 24 hrs. The slurries thus obtained were cast into thick films on silicon coated mylar sheets using a tape caster. It was possible to obtain tapes with the thickness in the range 10-40 $\mu$m. The tapes were arranged to obtain the desired multilayer (ML) structure, compacted under high pressure (3000 psi) and high temperature (400 K), and heated at 800-900 K for binder evaporation. The final sintering of the ML structure was carried out at 1300-1400 K. Samples consisted of (n + 1) layers of ferrite and n of layers of piezoelectric (n = 5 - 30). The thickness of layers varied from 10 to 40 microns.

Structural characterization was carried out using an x-ray diffractometer and CuK$_\alpha$ filtered radiation. In bulk samples, two sets of well-defined peaks were evident; the first set of narrow peaks corresponded to CZFO while the second set of relatively broad peaks was identified with the piezoelectric phase (PZT). The data did not show any detectable new (impurity) phases. But samples sintered at high temperatures (>1500 K) contained impurities. Structural parameters compared favorably with expected values for ferrites and PZT. Samples contained randomly oriented fine grains (1-5 micron) with a porosity of 5-10%.

X-ray diffraction for sintered multilayers did not show any impurities. Structural parameters for CZFO and PZT were calculated from the x-ray data. The lattice constant for CZFO increased linearly with Zn concentration and was in very good agreement with values for bulk materials [11]. The narrowness of the diffraction peaks and the estimated lattice parameters indicated that the structure was preserved and were free of interface strain. But the situation was somewhat different for PZT that showed several broad diffraction peaks. Bulk PZT was found to be tetragonal with $a$ = 0.4062(5) nm and $c$ = 0.4105(6) nm. But $a$-value was smaller for PZT in the composite and the reduction amounted to as much as 2.5% decrease in the unit cell volume of PZT. Thus x-ray data implies a strain free ferrite and a strained PZT in the heterostructures [7].

## 3. Bulk Composites

Techniques used for sample characterization included resistivity, dielectric constant, piezoelectric coupling constant, magnetization, magnetostriction and magnetoelectric voltage coefficient. Figure 1 shows the room temperature resistivity $\rho$ and the dielectric constant $\varepsilon$ (measured at 1 kHz) as a function of volume fraction $v$ for PZT in cobalt ferrite (CFO)-PZT. One observes six orders of magnitude increase $\rho$ and a factor of three increase in $\varepsilon$ as the PZT amount is increased from 0 to 100%. It is noteworthy that the dramatic increase in $\rho$ occurs for $v$>0.6, that coincides with the expected percolation limit for the ferrite in PZT. A similar behavior was evident for the piezoelectric coupling constant $d_{33}$. The magnetization was in agreement with expected values for CFO. For ME measurements, silver electrodes are deposited on thin disks of the composites and are polarized by heating it to 400 K and cooling it back to room temperature in an electric field (10-30 kV/cm) perpendicularly to the sample plane. Poled samples are placed between the pole pieces of an electromagnet that is fitted with a pair of Helmholtz coils. A static field H and an ac field $\delta H$=1 Oe (10

Hz – 1 kHz) are applied to the sample. The ME coefficient $\alpha'_E$ is determined by measuring the induced field $\delta E$ across the sample thickness. The measurements are done for two field orientations: (i) transverse or $\alpha'_{E,31}$ for H and $\delta H$ parallel to each other and to the disk plane (*1,2*) and perpendicular to $\delta E$ (direction-*3*) and (ii) longitudinal or $\alpha'_{E,33}$ for all the three fields parallel to each other and perpendicular to sample plane.

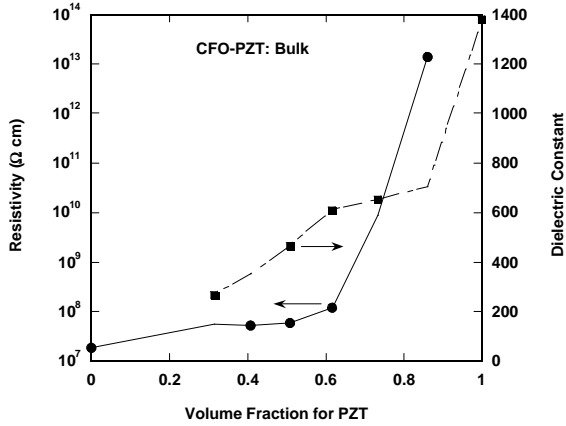

Fig.1: Room temperature electrical resistivity $\rho$ and dielectric constant $\varepsilon$ as a function of volume fraction $v$ for lead zirconate titanate (PZT) in bulk composites consisting of cobalt ferrite (CFO) and PZT.

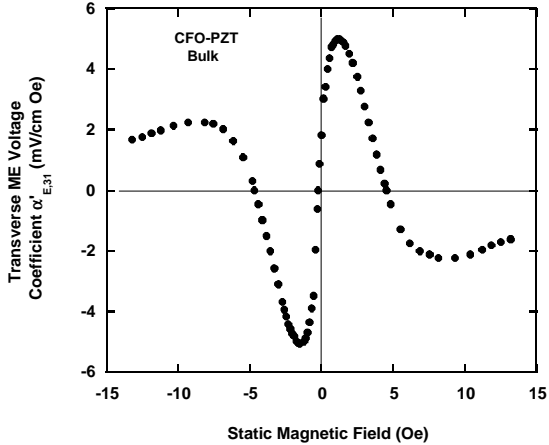

Fig.2: Transverse magnetoelectric (ME) voltage coefficient at 100 Hz as a function of bias magnetic field H for a bulk composite with PZT volume fraction $v = 0.4$. The transverse coefficient $\alpha'_{E,31} = \delta E_3/\delta H_1$ corresponds to H and an ac field $\delta H$ parallel to each other and to the sample plane (*1,2*) and the induced electric field $\delta E$ measured along the direction-*3*, perpendicular to sample plane.

Representative results on ME voltage coefficients and its dependence on H are shown in Fig.2 for a bulk composite with $v = 0.4$. The data for transverse field orientations are at room temperature for 100 Hz. One observes a general increase in $\alpha'_{E,31}$ with H to a peak value, followed by a rapid drop. With further increase in H, the coefficient drops to zero, reverses sign and then stays small. When H is reversed, one observes an additional phase difference of 180 deg. between $\delta H$ and $\delta E$. The longitudinal coefficient $\alpha'_{E,33}$ was of same magnitude as $\alpha'_{E,31}$ and exhibited similar H-dependence. The ME voltage coefficients are directly proportional to the product of pseudo-piezomagnetic coupling $q=\delta\lambda/\delta H$, where $\lambda$ is the magnetostriction, and the piezoelectric coefficient d. The H-dependence in Fig.2 essentially follows the slope of $\lambda$ vs H. When $\lambda$ attains saturation, the loss of piezomagnetic coupling leads to the absence of ME effect. Other features in Fig.2, such as zero-crossing and sign reversal, are due to competing qd terms and their variation with H [8-10].

Similar measurements of transverse and longitudinal voltage coefficients were done for CFO-PZT composites with the $v$ varying from 0 to 1. Figure 3 shows the maximum values of $\alpha'_{E,31}$ and $\alpha'_{E,33}$ as a function of volume fraction for PZT. The values were obtained from $\alpha'_E$ vs H as in Fig.2. The coefficients are zero in pure CFO or PZT; they increase as the PZT amount is increased. Key observations in Fig.3 are (i) a maximum in $\alpha'_{E,31}$ centered at $v = 0.75$ PZT and (ii) near-equal magnitudes for transverse and longitudinal coefficients. The coefficient $\alpha'_{E,31}$ is primarily due to transverse magnetostriction $\lambda_\perp$ whereas $\alpha'_{E,33}$ is due to the longitudinal magnetostriction $\lambda_{//}$. For a majority of ferrites $\lambda_{//}=2\ \lambda_\perp$ [11] and one expects $\alpha'_{E,33}=2\ \alpha'_{E,31}$. Such an empirical relationship is confirmed in samples in the shape of cubes [12]. The ratio of longitudinal to transverse coefficient in Fig.3, however, is quite small and ranges from 1.0 to 1.3. The deviation from the expected value of 2 is due to demagnetizing fields. There is no demagnetization effect for in-plane fields associated with the transverse ME effects. But the fields are applied perpendicular to the disk plane for the longitudinal case and demagnetization effects lead to a reduction in the internal ac and dc magnetic fields and a subsequent decrease in $\alpha'_E$ [13]. Similar studies were carried out on bulk composites of cobalt zinc ferrite and PZT. The ME voltage coefficients were quite small and the field and composition dependence of $\alpha'_E$ resembled data in Fig.3 and 4 for pure cobalt ferrite and PZT.

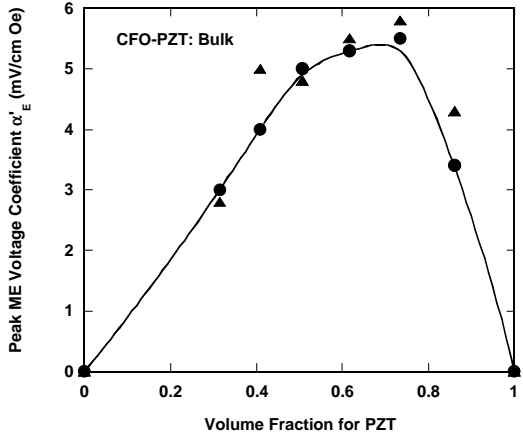

Fig.3: Dependence of maximum values of transverse (circles) and longitudinal (squares) ME voltage coefficients for bulk composites of CFO-PZT on volume fraction $v$ of PZT. The maximum values were obtained from data as in Fig.2. The longitudinal coefficient $\alpha'_{E,33} = \delta E_3/\delta H_3$ is for all the fields parallel to each other and along direction-$3$. The data are at room temperature and a frequency of 100 Hz. The lines are guide to the eye.

The most important inferences from Figs.2 and 3 are the weak ME interactions in bulk composites and the maximum in $\alpha'_E$ for PZT-rich composition. Theories for ME effects in bulk composites are based on specific particle shapes and mechanical connectivity between the two phases [4]. The voltage coefficients are directly proportional to the product of piezomagnetic and piezoelectric constants. Theories predict a maximum in $\alpha'_E$ for the composition with equal amount of ferrite and PZT. Estimates of $\alpha'_E$ based on, for example, cube shaped particles of $CoFe_2O_4$-$BaTiO_3$ and $CoFe_2O_4$-PZT range from 100 to 500 mV/cm Oe [4]. But the measured values are relatively small mainly due to low composite resistivity. The resistivity data in Fig.1 indicate orders of magnitude decrease in $\rho$ as the ferrite concentration is increased. The magnetic phase with a higher conductivity than PZT worsens conditions for polarization and weakens the piezoelectric and ME effects. Composite resistivity could be increase with substitutions such as Mn in the ferrite so that the formation of divalent Fe is eliminated [13]. Another option is the use of a layered structure and is considered next.

### 4. Layered Composites

The strongest ME coupling is expected in a layered structure due to (i) the absence of leakage current (ii) ease of poling to align the electric dipoles and strengthen the piezoelectric effect and (iii) control over mechanical and electrical connectivity between the two phases may be realized. Harshe, et al., proposed such structures, provided a theoretical model for a bilayer and prepared multilayers of CFO-PZT or $BaTiO_3$ by sintering thick films [4]. Our efforts on bilayers and multilayers (MLs) of CZFO-PZT are discussed here [7]. Samples studied contained CZFO with Zn concentration x=0 to 0.6. Measured magnetization was in agreement with expected values for bulk cobalt zinc ferrites [11]. Magnetostriction $\lambda$ was measured with a strain gage and a strain indicator. Figure 4 shows data on $\lambda$ measured parallel ($\lambda_{11}$) and perpendicular ($\lambda_{12}$) to the applied in-plane static field H for layered samples. The data at room temperature are shown for a series of Zn concentration and agree with values reported for bulk cobalt zinc ferrites [11]. The data are used later in this section for the determination of piezomagnetic coupling $q_{11}$ and $q_{12}$. These q-values are required for the determination of $\alpha'_{E,31}$.

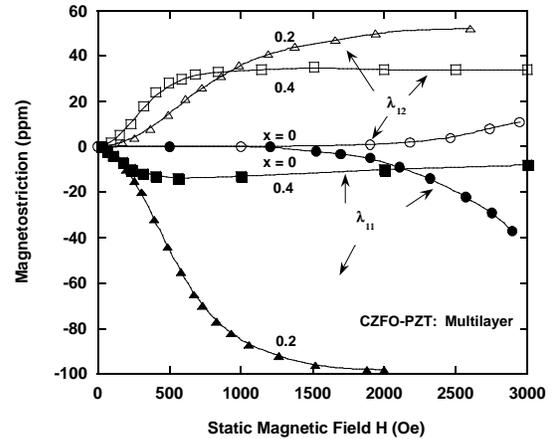

Fig.4: The magnetostriction $\lambda$ as a function of H for cobalt zinc ferrite, $Co_{1-x}Zn_xFe_2O_4$, with x=0, 0.2 and 0.4. $\lambda_{11}$ and $\lambda_{21}$ are in-plane parallel and perpendicular magnetostrictions, respectively.

Transverse and longitudinal ME voltage coefficients were measured on bilayers and multilayer samples of CZFO (x=0-0.6) and PZT. Representative data (at room temperature and 100 Hz) on H-dependence of ME coupling are shown in Fig.5 for $Co_{1-x}Zn_xFe_2O_4$-PZT fox x=0, 0.2 and 0.4.

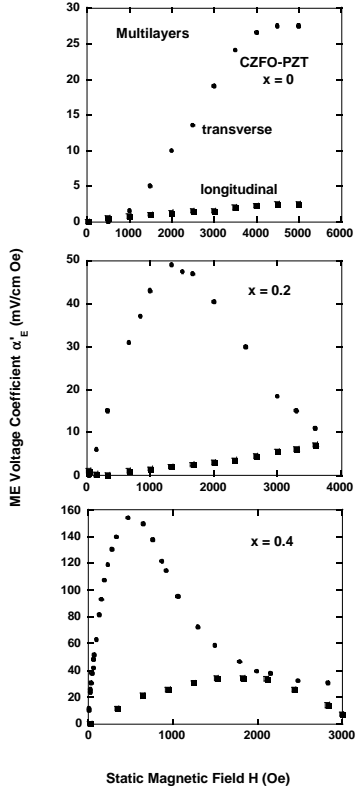

Fig.5: Variation of $\alpha'_{E,31}$ (circles) and $\alpha'_{E,33}$ (squares) with H at room temperature and 100 Hz for multilayer structures consisting of CZFO (x=0, 0.2 and 0.4) and PZT.

The $\alpha'_E$ values are for samples consisting of 16 ferrite and 15 PZT layers with a thickness of 20 µm. Consider first the data for pure cobalt ferrite (x=0)-PZT. As H is increased, $\alpha'_E$ increases with H to a maximum. The transverse coefficient is a factor of six higher than the longitudinal coupling and, as discussed later, is related to the relative strengths of piezomagnetic coupling coefficients. With Zn substitution for Co, one notices a pronounced increase in the magnitude of $\alpha'_E$. Data for x=0.2 indicate a factor of two increase in $\alpha'_E$ compared to CFO-PZT. Further increase in $\alpha'_E$ is evident for the sample with x=0.4. Other observations of importance include (i) the presence of longitudinal coupling over a wide H-range compared to the transverse case and (ii) down-shift in H-value corresponding to maximum in $\alpha'_E$ as x is increased. The overall H-dependence of $\alpha'_E$ in Fig.5 is similar in nature to results in Fig.2 for bulk composites. But the key observation in Fig.5 is the realization of the anticipated increase in the strength of $\alpha'_E$ compared to bulk composites (Fig.2 and 3). The transverse coefficient is a factor of 5-30 higher in multilayers than in the bulk. Since ME coupling originates from interactions at the ferrite-PZT interface, it is logical to associate strong ME coupling with a homogeneously mixed ferrite-PZT phases, as in a bulk composites. One expects a reduction in $\alpha'_E$ in layered systems due to a reduction in the interface area compared to bulk. But the results in Figs.2 and 5 are evidence for strong ME coupling in layered systems due to negligible leakage current, high piezoelectric coupling and control over mechanical and electrical connectivity.

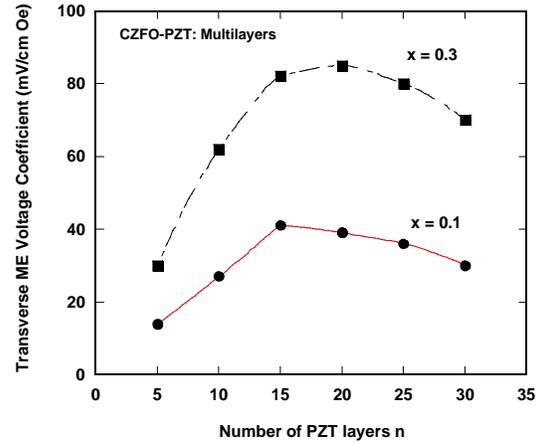

Fig.6: Dependence of peak transverse ME voltage coefficient on the number of PZT layers in multilayers of (n+1) layers of CZFO and n layers of PZT. The data are for Zn substitution x=0.1 and 0.3.

Our studies revealed a pronounced dependence of ME voltage coefficients on the number of layers. Figure 6 shows such data on $\alpha'_{E,31}$ vs n for CZFO (x=0.1 and 0.3)-PZT. The ME voltage increases initially with increasing n, then it stays constant for n=15-25 and shows a decrease for higher n. We measured the variation in maximum values of $\alpha'_E$ with Zn substitution in samples of CZFO (x=0-0.6)-PZT with n=15-25. The voltage coefficients were measured as a function of H and the peak values of $\alpha'_E$ thus obtained are shown in Fig.7 as a function of x. Measurements, when done for several samples, show variations in $\alpha'_E$ by as much as ±10% from the average value. Both the average value and the spread in $\alpha'_E$ are shown. The strength of ME interactions increase initially with increasing x. A maximum in $\alpha'_E$ is observed for x=0.4. Further increase in x leads to a decrease in $\alpha'_E$. The data show strong ME coupling in layered CZFO-PZT.

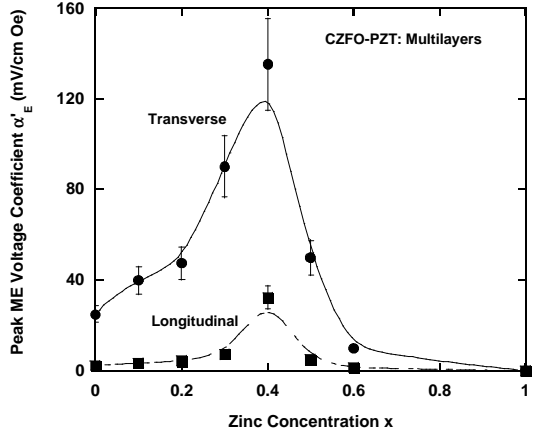

Fig.7: Dependence of maximum-$\alpha'_E$ on the Zn concentration in CZFO-PZT multiplayer composites. The bars represent the spread in measured values. The line is guide to the eye.

For comparison, the highest value for $\alpha'_E$ is 30 mV/cm Oe in bilayers of lanthanum manganites-PZT, 200 mV/cm Oe for NFO–PZT and 4680 mV/cm Oe for Terfenol-PZT [5,6,14].

A model for ME effects that assumes ideal coupling at the interface of a bilayer was developed in the past [4]. It did not consider the transverse fields for which one observes strong ME coupling and also ignored important effects related to finite magnetic permeability for the ferrite. We recently proposed a model for transverse and longitudinal ME effects that takes into account less-than-ideal interface coupling and the influence of demagnetization. We introduced a coupling parameter k to describe the interface coupling, with k=1 for an ideal interface and k=0 for a frictionless case [9,10]. An averaging method was used to obtain the following expressions for the ME voltage coefficients:

$$\alpha'_{E,33} = \frac{-2\mu_0 kv(1-v)\,^p d_{31}\,^m q_{31}}{2(^p d_{31})^2(1-v)k + ^p\varepsilon^T_{33}[(^p s_{11} + ^p s_{12})(v-1) - kv(^m s_{11} + ^m s_{12})]} \times$$

$$\times \frac{[(^p s_{11} + ^p s_{12})(v-1) - kv(^m s_{11} + ^m s_{12})]}{[\mu_0(v-1) - ^m\mu_{33} v][kv(^m s_{11} + ^m s_{12}) - (^p s_{11} + ^p s_{12})(v-1)] + 2(^m q_{31})^2 kv^2} \quad (1)$$

and

$$\alpha'_{E,31} = \frac{-kv(v-1)\,^p d_{31}(^m q_{11} + ^m q_{21})}{(^m s_{11} + ^m s_{12})^p\varepsilon^T_{33} kv + (^p s_{11} + ^p s_{12})\,^p\varepsilon^T_{33}(1-v) - 2(^p d_{31})^2 k(1-v)}. \quad (2)$$

Here p- and m- represent parameters for piezoelectric and magnetostrictive phases; $s$ and $d$ are compliance and piezoelectric coefficients; $\varepsilon$ is the permittivity at constant stress; $q$ and $\mu$ are piezomagnetic coefficients and the permeability; and $v = {^p v}/({^p v} + {^m v})$ with $^p v$ and $^m v$ denoting the volume of piezoelectric and magnetostrictive phases, respectively. Equations 1 and 2 are useful for the estimation of $H$ and $v$ dependence of ME coefficients.

Now we compare theoretical estimates for bilayers and data for multilayers. The following values were used for the material parameters: $^p s_{11} = 15*10^{-12}\ m^2/N$, $^p s_{12} = -5*10^{-12}\ m^2/N$, $^m s_{11} = 6.5*10^{-12}\ m^2/N$; $^m s_{12} = -2.4*10^{-12}\ m^2/N$, $d_{13} = -175\ pm/V$, $m_{33}/m_0 = 2$, and $\varepsilon_{33}/\varepsilon_0 = 1750$ [4, 13, 15]. For piezomagnetic coupling constants $q$, data on $\lambda$ vs $H$ were used. The magnetostriction $\lambda_{13}$ for CZFO is quite small and consequently $q_{13}$ is quite weak. The longitudinal coupling is, therefore, expected to be small and is in agreement with data in Fig.6 and 7. For transverse coupling, we first fitted $\lambda$ vs $H$ data in Fig.5 to a polynomial for the estimation of $q$ and its $H$ dependence. Equation 2 was then used to calculate $\alpha'_{E,31}$. Theoretical values for a bilayer of CZFO (x=0, 0.2 and 0.4)–PZT are compared with data in Fig.8. Results are for a sample with equal volume of the two phases ($v$=0.5). Estimated $\alpha'_{E,31}$ vs $H$ are shown as a function of the interface coupling constant $k$. For CFO-PZT (x=0), the theory predicts a gradual increase in $\alpha_{E,31}$ with increasing H. A maximum in $\alpha_{E,31}$ is expected for H = 3 kOe and the ME coefficient drops down to zero for H = 3.5 kOe, the field at which $(q_{11}+q_{12})$ vanishes. Above this field the effective $q$ is once again nonzero and gives rise to a ME voltage. Upon increasing x from 0 to 0.2, the theory predicts a sharp increase $\alpha_{E,31}$ at low fields, a down-shift in the peak position in $\alpha'_E$ to 500 Oe and a maximum $\alpha_{E,31}$ that is smaller than for x = 0. For x=0.4, the model predicts an intermediate minimum in $\alpha'_E$ at 200 Oe.

Next we compare the theory and data on $\alpha'_{E,31}$ for CZFO-PZT. Results of such comparison are shown in Fig.8 for x=0-0.4. For pure cobalt ferrite-PZT, we observe a substantial disagreement between theory and data. Neither the magnitude of $\alpha_{E,31}$ for $k$=1 nor its $H$ dependence agree with the data. The estimated $\alpha'_E$ are two orders of magnitude higher than data. We, therefore, infer a weak interface coupling between the ferrite and PZT layers with k on the order of 0.1. The coupling is somewhat enhanced with Zn substitution to k=0.2 for x=0.2. A significant increase in k, however, for x=0.4. Even though the data does not show the predicted intermediate minimum in $\alpha_{E,31}$, the estimated magnitude and H dependence for $k$=0.6 are in reasonable agreement with the data. Figure 9 shows the variation in the estimated $k$-values as a function of x. The coupling improves with the introduction of Zn, peaks for x=0.4 and decreases for higher x-values. A general improvement in the ME coupling is thus accomplished in layered composites with Zn substitution in the ferrite.

Now we discuss the possible cause of Zn-assisted enhancement in mechanical stress mediated electromagnetic coupling. Since this study is on the effects of magnetic parameters on ME coupling, one needs to examine critical ferrite parameters that control the magnetomechanical coupling. The dynamic piezomagnetic coupling in the ferrite arises due to magnetostriction and, in particular, the Joule magnetostriction caused by domain wall motion and domain rotation. Key requirements for strong coupling are unimpeded domain motion and a large $\lambda$. A soft, high initial permeability and high $\lambda$-ferrite is the main ingredient for giant ME effects. In magnetically hard cobalt ferrite, however, one has the disadvantage of a large anisotropy and coercive field that limits domain rotation. Enhancement in $\mu_i$ and $q$ could be accomplished with a single substitution, i.e., zinc in CFO. Figure 9 shows the variation in $\mu_i$ with x [11]. With the introduction of Zn, one observes an increase in $\mu_i$ to a maximum for x=0.5, followed by a decrease for higher x. It is clear from Fig.9 that as the Zn concentration is increased, both k and mi track each other and peak at x=0.4-0.5. The key inference here is the expected Zn-substitution assisted increase in the magneto-mechanical coupling and the consequent strengthening of ME interactions.

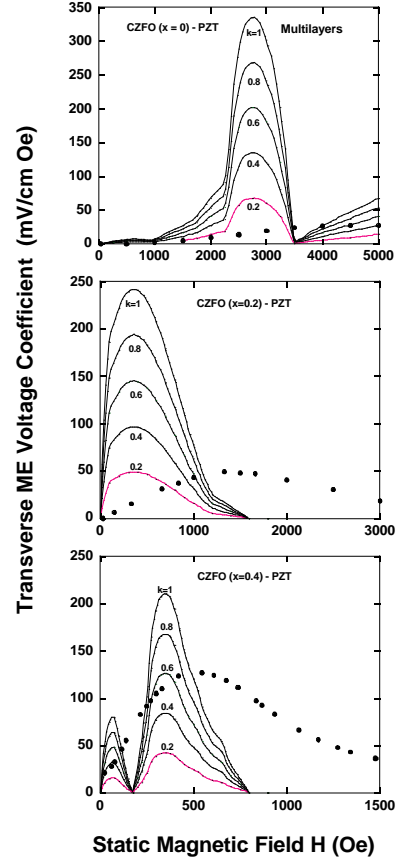

Fig.8: Comparison of theoretical and measured values of the transverse ME voltage coefficient $\alpha'_{E,31}$ for layered samples of CZFO (x=0, 0.2, 0.4) – PZT. The solid curves are theoretical values for a series for interface coupling parameter k.

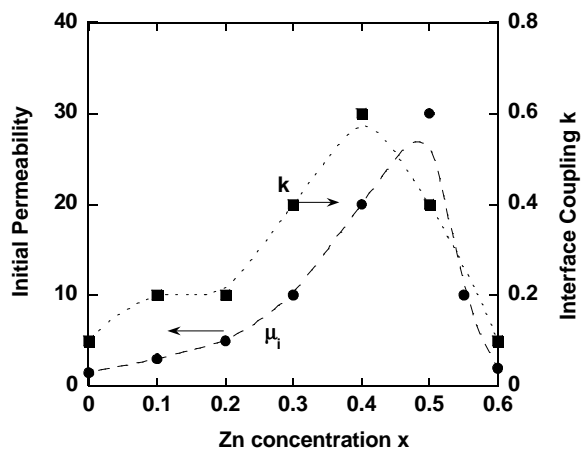

Fig.9: Composition dependence of the interface coupling constant k for CZFO-PZT multilayers obtained from comparison of theory and data as in Fig.8. The initial permeability $\mu_i$ (from Ref.11) is also shown as a function of x in CZFO and NZFO. The lines are guide to the eye.

## 5. Conclusion

We examined the ME coupling and its dependence on composition and static magnetic fields in bulk and layered samples of cobalt zinc ferrite and PZT. The coupling is rather weak in bulk samples, with equal magnitude for transverse and longitudinal fields. The low electrical resistivity for the ferrite is the primary cause of poor ME coupling. Such problems inherent to bulk samples could be eliminated in a layered structure. The strongest ME coupling for CZFO-PZT is realized in the layered structure due to the absence of leakage current and enhanced piezoelectric effect. The transverse coupling is the dominant ME interaction and is at least an order of magnitude higher than in the bulk samples. A distinct Zn-assisted increase in the ME coupling is evident in layered samples. Comparison of estimated H dependence of transverse coupling with data reveal a Zn-assisted improvement in the interface coupling. The improved coupling is attributed to efficient magneto-mechanical coupling in the ferrite.

**Acknowledgments**

The research was supported by a grant from the National Science Foundation (DMR-0302254).